\documentclass[12pt]{amsart}
\usepackage{amssymb,latexsym}

\title[Static and Dynamic traversable wormhole geometries]
{Static and dynamic traversable wormhole geometries satisfying the Ford-Roman constraints}
\author{Peter K.F. Kuhfittig}
\address{Department of Mathematics\\
Milwaukee School of Engineering\\
Milwaukee, Wisconsin 53202-3109}
\date{\today}

\begin{document}

\maketitle

\begin{abstract}
It was shown by Ford and Roman in 1996 that quantum field theory severely constrains wormhole geometries on a
macroscopic scale.  The first part of this paper discusses a wide class of wormhole solutions
that meet these constraints.  The type of shape function used is essentially
generic.  The constraints are then discussed in conjunction with various redshift
functions.  Violations of the weak energy condition and traversability 
criteria are also considered.  

The second part of the paper analyzes analogous time-dependent
(dynamic) wormholes with the aid of differential forms.  It is shown that a violation
of the weak energy condition is not likely to be avoidable
even temporarily.

PACS number(s): 04.20.Jb, 04.20.Gz
\end{abstract}

\section{Introduction}
Wormholes may be defined as handles or tunnels in the spacetime topology linking widely separated regions 
of our universe or even of different universes.  The possible existence of wormholes was
already recognized by Ludwig Flamm in 1916~\cite{lF16}.  The now classic work of Einstein and
Rosen~\cite{aE35} dates from 1935, while the work of Wheeler~\cite{jW55} in the 1950's initiated
the modern era.  More recently, the discovery of traversable wormholes by Morris and
Thorne~\cite{MT88,MTY88} resulted in a flurry of activity that continues to the present
and has led to speculations about interstellar travel and even time travel \cite{MTY88}.
An excellent survey of these developments can be found in the book by Visser~\cite{mV96}.  

As Morris and Thorne ~\cite{MT88} observed, to hold a wormhole open, violations of certain energy conditions
must be tolerated.  More precisely, all known forms of matter obey the weak energy condition
(WEC) $T_{{\alpha}{\beta}}{\mu}^{\alpha}{\mu}^{\beta}\ge0$ for all time-like vectors and,
by continuity, all null vectors.  For a detailed discussion see Friedman ~\cite{jF97}.  If
the WEC is violated, then the energy density of matter may be seen as negative by some
observers.  Morris and Thorne call such matter ``exotic."

A detailed analysis by Ford and Roman ~\cite{lF96} in 1996 showed that quantum field theory constrains the 
wormhole geometries so that some of the earlier wormhole solutions proved to be problematic on a
macroscopic scale: either the wormhole is only slightly larger than Planck size or there
exist large discrepencies in the length scales that describe the wormhole.  At best, the
negative energy matter will have to be confined to a shell very much thinner than the
throat.  Various solutions meeting these constraints are discussed in Ref. \cite{MTY88}, \cite
{pK99}, and \cite{aD01}.

In this paper we treat these restrictions as basic requirements for any (traversable) wormhole.  
A wide
class of wormhole solutions can nevertheless be constructed.  In particular, the shape
functions are quite general, essentially generic.  The constraints are then discussed
in conjunction with different redshift functions.  As usual, the wormholes are assumed to be
spherically symmetric, joining distant regions that are asymptotically flat.  The last
requirement is needed in the discussion of wormhole size and traversability conditions.

This paper is divided into two parts.  In the first part (Section \ref{S:timeindependent}) 
the time-independent solutions are
discussed.  This allows a straight-forward analysis of the weak energy condition
in terms of the redshift and shape functions.  Most of the discussion of traversability
conditions assumes time-independence.

The second part (Section \ref{S:timedependent}) discusses analogous time-dependent (dynamic)
wormholes using a very general form of the line element.  It is shown
that even a temporary suspension of the weak energy condition
is likely to yield an event horizon, in which case one no longer
has a wormhole, or the tidal forces increase until the wormhole
is no longer traversable.  This section uses the method of
differential forms and is largely independent of the first part of the paper.

\section{Time-Independent Solutions}\label{S:timeindependent}
   \subsection{The Shape Function}\label{SS:shape}
Our starting point is the spherically symmetric line element 
\begin{equation}\label{E:line1}
   ds^2=-e^{2\gamma(r)_\pm}c^2dt^2+e^{2\alpha(r)_\pm}dr^2+r^2(d\theta^2
   +\text{sin}^2\theta
   \,d{\phi}^2)
\end{equation}
where $-\infty<t<\infty,\,\, 0<r_0<r<\infty,\,\, 0<\theta<\pi, \,\text{and}\,\,\,0\leq\phi<2\pi.$
Here $\gamma(r)_{\pm}$ and $\alpha(r)_{\pm}$ are functions of the radial coordinate $r$.  The
subscript $\pm$ refers to the respective ``upper" and ``lower" universes but will not be
used in the remainder of the paper.  (We also refrain from adopting units in which $c=1$
to facilitate the calculations in Sections \ref{SS:wormsize} and \ref{SS:traversability}.)

In line element~(\ref{E:line1}), the functions $\gamma(r)$ and $\alpha(r)$ can be
freely assigned to yield the desired wormhole properties.  For the former,
called the \emph{redshift function}, we demand that $e^{2\gamma (r)}$ not have
any zeros; in other words, there is to be no event horizon.  The function
$\alpha (r)$ determines the \emph{shape function} $b=b(r)$, defined below.

The wormhole geometry is usually described by means of an embedding diagram in
three-dimensional Euclidean space at a fixed moment in time and for a fixed
value of $\theta$, the equatorial slice $\theta=\pi/2$ (Ref.~\cite{MT88, aD01}).
The resulting surface of revolution has the parametric form
\begin{equation}\label{E:parametric}
   f(r,\phi)=(r\,\text{cos}\,\phi,\,r\,\text{sin}\,\phi,\,z(r))
\end{equation}
for some function $z=z(r)$.  This surface connects the
asymptotically flat upper and lower universes.  The radial coordinate decreases
from $+\infty$ in the upper universe to a minimum value $r=r_0$ corresponding
to the throat of the wormhole, and then increases again from $r_0$ to
$+\infty$ in the lower universe.

The function $z=z(r)$ in Eq.~(\ref{E:parametric}) must have the 
following properties:
\begin{equation}\label{E:embeddingflare}
   \lim_{r \to \infty} \frac{dz}{dr}=0,
\end{equation}
the meaning of asymptotic flatness. 
At the throat $r=r_0$,  
\begin{equation}\label{E:embeddingthroat}
   \lim_{r \to r_0+} \frac{dz}{dr}=+\infty.
\end{equation}
(The embedding surface must have a vertical tangent at the throat.)
Returning to the line element~(\ref{E:line1}), we further assume that 
$\alpha(r)$ has a vertical asymptote at $r=r_0$: 
$\lim_{r \to r_0+}\alpha(r)=+\infty$.  In addition, $\alpha(r)$ is
twice differentiable and strictly decreasing with $\lim_{r \to \infty}\alpha(r)=0$.

It is now seen that these requirements are met by $z=z(r)$ such that
\begin{equation}\label{E:embeddingshape}
   \frac{dz}{dr}=\sqrt{e^{2\alpha(r)}-1}
\end{equation}
(for the upper universe).  Moreover, $d^2z/dr^2<0$ near the throat (since
$\alpha'(r)<0)$, as required by the ``flaring out" condition (Ref.~\cite{MT88}).

The shape function $b=b(r)$ is now defined by
\[
    e^{2\alpha(r)}=\frac{1}{1-\frac{b(r)}{r}}.
\]
The shape function determines the spatial shape of the wormhole as viewed 
in an embedding diagram.  For example, the asymptotic flatness may now
be described by the condition
\[
   \lim_{r \to \infty}\frac{b(r)}{r}=0.
\]
At the throat we have $b(r_0)=r_0$, as in the case of a Schwarzschild wormhole.

To obtain the general form of the shape function $b=b(r)$, we 
return to the parametric form~(\ref{E:parametric}),
\[
  f(r,\phi)=(r\,\text{cos}\,\phi,\,r\,\text{sin}\,\phi,\,z(r)).
\]
To determine the induced metric $ds_1^2=dx^2+dy^2+dz^2$ on this surface, compute the three
differentials and substitute, yielding $ds_1^2=dr^2+r^2\,d\phi^2+(z'(r))^2\,dr^2$, or
\[
  ds_1^2=[1+(z'(r))^2]\,dr^2+r^2\,d\phi^2.
\]
  It follows from
\[
  1+(z'(r))^2=\frac{1}{1-\frac{b(r)}{r}}
\]
and $dz/dr=\sqrt{e^{2\alpha(r)}-1}$ that
\begin{equation}\label{E:shape function}
   b(r)=r(1-e^{-2\alpha(r)}).
\end{equation}
This is the general form of the shape function.  We assume only that $\alpha(r)$ is twice
differentiable for $r>r_0$.  (It will be seen later that for physical reasons
$\alpha(r)$ must ``level off" sharply.)

   \subsection{The Redshift Function}
Since we are using a very general form for the shape function, a suitable form
for the redshift function needs to be found.  An excellent choice is $
\gamma(r)=-\kappa/r$, proposed by Anchordoqui \emph{et al.}\,\,\cite{lA98}.  Here
$\kappa$ is a positive constant to be determined later.  Since $r=0$ is not part of the manifold,
there is no event horizon.  This choice is particularly convenient for determining the size
of the wormhole, as well as the traversability conditions, in 
Sections \ref{SS:wormsize}
 and \ref{SS:traversability}.  

Another, more general, way to avoid an event horizon is to denote the function $\alpha$ by
 $\alpha(r-r_0)$ to
emphasize its behavior at the throat and to use the translated curve $\gamma(r)=-\alpha(r)$
for the redshift function.  Thus $\lim_{r \to 0+}\alpha(r)=\infty$.  The line element then becomes
\begin{equation}\label{E:line2}
   ds^2=-e^{-2\alpha(r)}c^2dt^2+e^{2\alpha(r-r_0)}dr^2
   +r^2(d\theta^2+\text{sin}^2\theta
   \,\,d\phi^2).
\end{equation}
In Section \ref{SS:redshift} the solution will be extended to the time-dependent redshift function $\gamma(r)
=-\lambda(r,t).$

To study the traversability conditions, we need the components of the Riemann curvature tensor.
As will become apparent later, making the redshift function time-dependent
does not result in any additional nonzero components.  The calculations
have therefore been carried out for $\gamma(r)=-\lambda(r,t)$,
 this case being of particular interest.  The results are
given in the Appendix.  The special
cases $\gamma(r)=-\kappa/r$ and $\gamma(r)=-\alpha(r)$ follow.  In fact, for 
$\gamma(r)=-\alpha(r)$, we have in the following nonzero components expressed in the
orthonormal frame:
\begin{multline*}
    R_{\hat{r}\hat{t}\hat{r}}^{\hat{t}}
    =-R_{\hat{r}\hat{r}\hat{t}}^{\hat{t}}
   =R_{\hat{t}\hat{t}\hat{r}}^{\hat{r}}
    =-R_{\hat{t}\hat{r}\hat{t}}^{\hat{r}}\\
      =e^{-2\alpha(r-r_0)}\left[\alpha''(r)
     -\alpha'(r)\alpha'(r-r_0)
      -(\alpha'(r))^2\right],
\end{multline*}
\begin{equation*}
   R_{\hat{\theta}\hat{t}\hat{\theta}}^{\hat{t}}
   =-R_{\hat{\theta}\hat{\theta}\hat{t}}^{\hat{t}}
   =R_{\hat{t}\hat{t}\hat{\theta}}^{\hat{\theta}}
   =-R_{\hat{t}\hat{\theta}\hat{t}}^{\hat{\theta}}=
   \frac{1}{r}e^{-2\alpha(r-r_0)}\alpha'(r),
\end{equation*}
\begin{equation}\label{E:Riemann}
   R_{\hat{\phi}\hat{t}\hat{\phi}}^{\hat{t}}
   =-R_{\hat{\phi}\hat{\phi}\hat{t}}^{\hat{t}}
   =R_{\hat{t}\hat{t}\hat{\phi}}^{\hat{\phi}}
   =-R_{\hat{t}\hat{\phi}\hat{t}}^{\hat{\phi}}
   =\frac{1}{r}e^{-2\alpha(r-r_0)}\alpha'(r),
\end{equation}
\begin{equation*}
   R_{\hat{\theta}\hat{r}\hat{\theta}}^{\hat{r}}=
   -R_{\hat{\theta}\hat{\theta}\hat{r}}^{\hat{r}}
   =R_{\hat{r}\hat{\theta}\hat{r}}^{\hat{\theta}}
   =-R_{\hat{r}\hat{r}\hat{\theta}}^{\hat{\theta}}
   =\frac{1}{r}e^{-2\alpha(r-r_0)}\alpha'(r-r_0),
\end{equation*}
\begin{equation*}
   R_{\hat{\phi}\hat{r}\hat{\phi}}^{\hat{r}}
   =-R_{\hat{\phi}\hat{\phi}\hat{r}}^{\hat{r}}
   =R_{\hat{r}\hat{\phi}\hat{r}}^{\hat{\phi}}
   =-R_{\hat{r}\hat{r}\hat{\phi}}^{\hat{\phi}}
   =\frac{1}{r}e^{-2\alpha(r-r_0)}\alpha'(r-r_0),
\end{equation*}
\begin{equation*}
   R_{\hat{\phi}\hat{\theta}\hat{\phi}}^{\hat{\theta}}
   =-R_{\hat{\phi}\hat{\phi}\hat{\theta}}^{\hat{\theta}}
   =R_{\hat{\theta}\hat{\phi}\hat{\theta}}^{\hat{\phi}}
   =-R_{\hat{\theta}\hat{\theta}\hat{\phi}}^{\hat{\phi}}
   =\frac{1}{r^2}\left(1-e^{-2\alpha(r-r_0)}\right).
\end{equation*}

The components of the Ricci tensor are also listed in the Appendix.  The components
of the Einstein tensor are given by (from $G_{\hat{\alpha}\hat{\beta}}=
R_{\hat{\alpha}\hat{\beta}}-\frac{1}{2}Rg_{\hat{\alpha}\hat{\beta}})$
\begin{equation*}
   G_{\hat{t}\hat{t}}=\frac{2}{r}e^{-2\alpha(r-r_0)}\alpha'(r-r_0)+
   \frac{1}{r^2}\left(1-e^{-2\alpha(r-r_0)}\right),
\end{equation*}
\begin{equation}\label{E:Einstein}
   G_{\hat{r}\hat{r}}=-\frac{2}{r}e^{-2\alpha(r-r_0)}\alpha'(r)-
   \frac{1}{r^2}\left(1-e^{-2\alpha(r-r_0)}\right),
\end{equation}
\begin{multline*}
  G_{\hat{\theta}\hat{\theta}}=G_{\hat{\phi}\hat{\phi}}
  =e^{-2\alpha(r-r_0)}\bigg[\quad\negthickspace\negthickspace
     \negthickspace\negthickspace
   -\alpha''(r)
   +\alpha'(r)\alpha'(r-r_0)
   +(\alpha'(r))^2\\
       \left.-\frac{1}{r}\alpha'(r)
       -\frac{1}{r}\alpha'(r-r_0)\right].
\end{multline*}

\subsection{Analyzing the WEC violation}
We now show that near the throat the weak energy condition is violated.  In
the orthonormal frame the basis vectors are those used by static observers.  As
a result, the components of the stress-energy tensor are simply
\begin{equation}\label{E:stress}
   T_{\hat{t}\hat{t}}=\rho c^2,\, \,T_{\hat{r}\hat{r}}=-\tau,\,\, 
   \text{and}\,\,\, 
   T_{\hat{\theta}\hat{\theta}}=T_{\hat{\phi}\hat{\phi}}=p,
\end{equation}
all of which are functions of the radial coordinate $r$ for static wormholes.

Consider next the weak energy condition $T_{\hat{\alpha}\hat{\beta}}\mu^{\hat{\alpha}}\mu^{\hat{\beta}}
\ge0$, where $\mu^{\hat{\alpha}}=(\mu^{\hat{t}},\mu^{\hat{r}},0,0)=(1,1,0,0)$
is a radial outgoing null vector.  The condition now becomes
$T_{\hat{t}\hat{t}}+T_{\hat{r}\hat{r}}=\rho c^2-\tau\ge0$.
Given the Einstein field equations
\begin{equation}\label{E:field eq.}
   G_{\hat{\alpha}\hat{\beta}}=8\pi Gc^{-4}T_{\hat{\alpha}\hat{\beta}}
\end{equation}
we evidently need to examine
\begin{equation}\label{E:exotic1}
   \rho c^2-\tau=T_{\hat{t}\hat{t}}+T_{\hat{r}\hat{r}}=
   \frac{1}{8\pi Gc^{-4}}\left[\frac{2}{r}e^{-2\alpha(r-r_0)}
   \left(\alpha'(r-r_0)
   -\alpha'(r)
    \right)
    \right].
\end{equation}
This equation shows that $\rho c^2-\tau\ < 0$ near the throat (but not necessarily
at the throat).  Moreover, if $\alpha(r-r_0)$ is steep enough (i.e., if
$|\alpha'(r-r_0)|$ decreases fast enough), then the interval containing exotic
matter can be made as small as necessary to satisfy the quantum inequality of
Ford and Roman ~\cite{lF96}.  This can be seen more easily with the redshift function
$\gamma(r)=-\kappa/r$ in line element (\ref{E:line1}):
\begin{equation}\label{E:exotic2}
   \rho c^2-\tau=\frac{1}{8\pi Gc^{-4}}\left[\frac{2}{r}
   e^{-2\alpha(r-r_0)}
   \left(\alpha'(r-r_0)
   +\frac{\kappa}{r^2}\right)\right].
\end{equation}
In other words, up to now $\alpha(r-r_0)$ was assumed to be any twice 
differentiable function approaching the asymptote $r=r_0$ 
from the right, but now we see that for physical reasons
some restriction must be placed on its slope:
choose $\alpha(r-r_0)$ so that $|\alpha'(r-r_0)|\leq\kappa/r^2$
outside the interval $[r_0,r_1]$ for some $r_1$.

With these assumptions, $b'(r)$ is close to unity near the throat, causing the embedding function to flare
out very slowly, also in agreement with \cite{lF96}.

\section{Traversability Conditions}
\subsection{Wormhole Size}\label{SS:wormsize}
To estimate the size of the wormhole and to check the traversability conditions, we need
to rely on specific examples.

One solution that meets the constraints mentioned earlier is the Morris-Thorne-Yurtsever
wormhole (MTY) discussed in Ref.~\cite{MTY88}. Particularly encouraging is the
fact that this wormhole makes use of the experimentally confirmed Casimir
effect, that is, the effect produced by two parallel conducting plates
a small distance apart, resulting in a violation of the weak energy condition.
Accordingly, the MTY wormhole 
consists of a pair of spherical
charged Casimir plates within a very small proper distance of each other, positioned
on each side of the throat.  The energy density is negative between the plates and near
exotic outside the plates, i.e., the outside is a classical radial Coulomb field
with $\rho c^2=\tau=p=Q^2/8\pi r^4$, which produces a Reissner-Nordstrom spacetime
geometry \cite{MTY88}.

As noted earlier, since $b'(r)$ is likely to be close to unity near the throat,
the wormhole will flare out very slowly.  Consequently, $r_0$ is going to be
relatively small; so we may assume that $r_0\approx0$.
As a result, referring to the line element (\ref{E:line1}), we may choose 
\begin{equation}\label{E:estimates}
  \gamma(r)=-\frac{\kappa}{r}=-\frac{0.00025}{r}\,\,\, \text{and} \,\, \,\alpha(r)
  =\frac{0.00025}{r}.
\end{equation}
Then $b(r)=r(1-e^{-0.0005/r}$).

To obtain the desired estimates, we consider the thought experiment in
Morris and Thorne (Ref.~\cite{MT88}).  They assumed that the spaceship 
travels radially through the wormhole starting at a space station located 
at a safe distance from the wormhole.  In particular,  
 we would like the station to be
far enough away from the throat so that $1-b(r)/r$ is within 1\% of unity:
\[
  1-\frac{b(r)}{r}=e^{-0.0005/r}\approx0.99
\]
yielding $r=0.0497$ l.y.  As in Ref.~\cite{MT88}, we assume that the spaceship
accelerates at $g_\oplus=9.8$ m/$\text{s}^2$ halfway to the throat and then
decelerates at the same rate until it comes to rest near the throat.  We have
$s=\frac{1}{2}[0.0497\times9.46\times10^{15}$ m] and $t=\sqrt{s/(g_\oplus/2)}=
6~926~450$ s$\,\,\approx $\,\,80 days.  The total length of the trip, then, is less
than one year, as recommended in Ref.~\cite{MT88}.  (The maximum velocity is about
68~000 km/s.)

\subsection{Other Traversability Conditions}\label{SS:traversability}
The choice of $\kappa=0.00025$ not only yields the relatively low travel time of 80 days
for the first part of the trip, it also helps in satisfying conditions (40) and (43) in
Ref.~\cite{MT88}.  For example, at the space stations condition (40) requires that
$b/r\ll1$ and $|\,\Phi|\ll1$, where $\Phi$ is the redshift function.  We have
\[
  \frac{b}{r}=1-e^{-0.0005/r}\approx 0.01\ll1 \,\,\text{and}\,\,
  |\,\Phi|=\frac{\kappa}{r}\approx0.005\ll1.
\]
The redshift function must also satisfy, again by condition (40),
\[
  \Phi'\leq\frac{g}{c^2\sqrt{1-b/r}}.
\]
At the station, $b/r\approx 0$, so the condition becomes
$\Phi'\leq(9.2\times10^{15}\,\text{m})^{-1}$.  Since $\kappa$ has units of length,
we have indeed
\[
  \Phi'=\frac{\kappa}{r^2}=\frac{0.00025\times9.46
   \times10^{15}\,\text{m}}
  {(0.0497\times9.46\times10^{15}\,\text{m})^2}\approx 1.1\times10^{-17}
  \,\text{m}^{-1}.
\]
For the MTY wormhole, $r_0=1$ A.U.$\,\approx\,0.0000158$ l.y.  So near the throat,
\[
  \Phi'\approx\frac{0.00025\times9.46\times10^{15}\,\text{m}}{(0.0000158
  \times9.46\times10^{15}\,\text{m})^2}\approx 1.1
  \times10^{-10}\,\text{m}^{-1},
\]
which is less than
\[
  \frac{g}{c^2\sqrt{1-b/r}}=\frac{1}{9.2\times 10^{15}\sqrt
  {1.8\times 10^{-14}}}\approx 8.1\times 10^{-10}\,\text{m}^{-1}.
\]
Condition (43), $|\,a|\leq g_\oplus/c^2$, where
\[
  a=\pm\left(1-\frac{b}{r}\right)^{\frac{1}{2}}e^{-\Phi}\left(\gamma
  e^{\Phi}\right)'c^2,
\]
which is a constraint on the traveler's acceleration, follows directly 
since the Lorentz parameter\,\,$\gamma$\, is close to unity.

To analyze the tidal gravitational forces that an infalling radial observer may feel during the
journey, we need to find some of the components of the Riemann curvature tensor
relative to the following orthonormal basis (from the usual Lorentz transformation):

\begin{align*}
   e_{\hat{0}'}&=\gamma e_{\hat{t}}\mp\gamma\left(\frac{v}{c}\right)e_{\hat{r}},&
   e_{\hat{1}'}&=\mp\gamma e_{\hat{r}}+\gamma\left(\frac{v}{c}\right)e_{\hat{t}},\\
       e_{\hat{2}'}&=e_{\hat{\theta}}, & e_{\hat{3}'}&=e_{\hat{\phi}}.\end{align*}
As suggested in Ref.~\cite{MT88}, the tidal forces that an observer would feel 
must not exceed the ones experienced on Earth.  As outlined in
Ref.~\cite{MT88}, the radial tidal constraint is given by
\begin{equation*}
   \left|R_{\hat{1}'\hat{0}'\hat{1}'\hat{0}'}\right|\leq
    \frac{g_\oplus}{c^2\times 2\,\text{m}}\approx
     \frac{1}{(10^8\,\text{m})^2},
\end{equation*}
assuming an observer $2\,\text{m}$ tall.  Thus, 
from Equations~(\ref{E:Riemann}) and
(\ref{E:estimates}), 
\begin{multline}\label{E:radialtidal}
  \left|R_{\hat{1}'\hat{0}'\hat{1}'\hat{0}'}\right|
  =\left|R_{\hat{r}\hat{t}\hat{r}\hat{t}}\right|
  =\left|e^{-2\kappa/r}\left[\frac{2\kappa}{r^3}
  +\frac{\kappa}{r^2}\left(-\frac{\kappa}{r^2}\right)
  -\frac{\kappa^2}{r^4}\right]\right|\\
        =\left|e^{-0.0005/r}\left(\frac{0.0005}{r^3}-
        \frac{2(0.00025)^2}{r^4}\right)\right|,
\end{multline}
which is well below $(10^8\,\text{m})^{-2}$ for all $r$ in the interval $(0,\infty)$.

For the lateral tidal force we have
\begin{equation}\label{E:lateraltidal}
   \left|R_{\hat{2}'\hat{0}'\hat{2}'\hat{0}'}\right|
   =\left|R_{\hat{3}'\hat{0}'\hat{3}'\hat{0}'}\right|=\gamma^2\left|
   R_{\hat{\theta}\hat{t}\hat{\theta}\hat{t}}\right|
         +\gamma^2\left(\frac{v}{c}\right)^2\left|
         R_{\hat{\theta}\hat{r}\hat{\theta}\hat{r}}\right|,                
\end{equation}
which also meets the constraint 
$|R_{\hat{2}'\hat{0}'\hat{2}'\hat{0}'}|\leq(10^8\,\text{m})^{-2}$
even at high velocities.  

\section{Time-dependent solutions}\label{S:timedependent}
\subsection{Introduction}
A number of studies in recent years have dealt with time-dependent (dynamic) wormhole 
solutions.  A particularly interesting example is the possibility of
enlarging a submicroscopic wormhole to macroscopic size considered by
Roman~\cite{tR93} using the line element
\begin{equation*}
   ds^2=-e^{2\Phi(r)}dt^2\\
      +e^{2\chi t}\left[\frac{dr^2}{1-b(r)/r}+
      r^2(d\theta^2+\sin^2\theta \,d\phi^2) \right].
\end{equation*}
Kim ~\cite{swK96} generalized the Morris-Thorne wormhole by using a scale
factor $R(t)$:
\begin{equation*}
   ds^2=-e^{2\Phi(r)}dt^2+R^2(t)\left[\frac{dr^2}{1-kr^2-b(r)/r}+r^2
   (d\theta^2+\sin^2\theta\,d\phi^2)\right].
\end{equation*}
It was concluded in the former case that a violation of the WEC
cannot be avoided, while in the latter case it can.

Line elements with a conformal factor $\Omega(t)$, that is, 
\begin{equation*}
   ds^2=\Omega(t)\left[-e^{2\Phi(r)}dt^2+e^{2\Lambda(r)}dr^2+r^2
   (d\theta^2+\sin^2\theta\,d\phi^2)\right]
\end{equation*}
were considered by Anchordoqui \emph{et al.}\,\,\cite{lA98},
Kar~ \cite{sK94}, and Kar and Sahdev ~\cite{KS96}.  All have
concluded that the WEC violation can be avoided for various
intervals of time.  Similar conclusions can be found in \cite{WL95}.

In this paper we analyze the more natural case in which both $\gamma$
and $\alpha$ in line element ~(\ref{E:line1}) are time-dependent,
thereby differentiating our results from those cited above.
This approach clearly points out the consequences of trying to
eliminate the exotic matter even temporarily.

\subsection{The Time-Dependent Redshift Function}\label{SS:redshift}
Suppose the red-shift function is time-dependent, i.e.,
$\gamma(r)=-\lambda(r,t)$.  Then the weak energy condition becomes $\rho(r)c^2-\tau(r,t)
\ge0$.  So we need to examine the following equation, using the information in the Appendix:
\begin{multline}\label{E:exotic3}
   \rho(r)c^2-\tau(r,t)\\=\frac{1}{8\pi Gc^{-4}}\left[\frac{2}{r}
   e^{-2\alpha(r-r_0)}\left(\alpha'(r-r_0)-
   \frac{\partial}{\partial r}\lambda(r,t)\right)\right].
\end{multline}
Since $\alpha'(r-r_0)$ is still the same, this condition is 
similar to Condition~(\ref{E:exotic1}) for any fixed $t$. There is one important
difference, however: since $\lambda(r,t)$ varies with time, there may exist time 
intervals in which
\[
  \alpha'-\frac{\partial}{\partial r}\lambda(r,t)=0
\]
in the vicinity of the throat.  This suggests that the weak energy condition 
 and the need to use exotic matter can be temporarily
 suspended.  Unfortunately, in this time interval, $\lambda(r,t)=\alpha+k$, where
$k$ is a finite quantity.  So the first term in the line element ~(\ref{E:line1}) 
becomes zero at the throat, thereby creating an event horizon.  A similar 
conclusion was reached by Hochberg and Visser~\cite{dH98} by defining 
a wormhole throat to be a marginally anti-trapped surface.  For
a discussion of whether this definition necessarily applies to all
wormholes, see Li~\cite{lxL01}.  Our conclusions hold for the type
of wormhole in the studies cited above.

\subsection{The Time-Dependent Function $\alpha(r,t)$}
If $\alpha(r,t)$ and the resulting shape function $b=r(1-e^{-2\alpha(r,t)})$
are also time-dependent, we will get additional nonzero components of
the Riemann curvature tensor.  As a result, the calculations in the Appendix
can no longer be used.  The safest way to obtain these components
is by the use of differential forms (Cartan's method).  To do so,
we will follow the book by Hughston and Tod ~\cite{HT90}.  (The
remainder of this paper is independent of the preceding part.)

The line element is
\begin{equation}\label{E:line3}
   ds^2=-e^{-2\lambda(r,t)}c^2 dt^2+e^{2\alpha(r,t)} dr^2
   +r^2 d\theta^2+r^2\,\text{sin}^2\theta\, d\phi^2,
\end{equation}
where both $\lambda$ and $\alpha$ are time-dependent.  Given the basis
\[
   \theta^0=e^{-\lambda(r,t)}dt,\thickspace \theta^1=e^{\alpha(r,t)}dr,
      \thickspace\theta^2=r\,d\theta, 
      \thickspace\theta^3=r\,\text{sin}\,\theta\,d\phi,
\]
we calculate the exterior derivatives:
\[
    d\theta^0=-\frac{\partial}{\partial r}\lambda(r,t)
    e^{-\alpha(r,t)}\,\theta^1\wedge \theta^0,
 \]
 \[
     d\theta^1=\frac{\partial}{\partial t}\alpha(r,t)
     e^{\lambda(r,t)}\,\theta^0\wedge \theta^1,
\]
\[
     d\theta^2=\frac{1}{r}e^{-\alpha(r,t)}\,\theta^1\wedge\theta^2,
\]
\[
     d\theta^3=\frac{1}{r}e^{-\alpha(r,t)}\,\theta^1\wedge\theta^3
      +\frac{1}{r}(\text{cot}\,\theta)\,\,\theta^2\wedge\theta^3.
\]
The connection 1-forms $\omega^i_{\phantom{i}\,\,k}$ have the symmetry
\[
    \omega^0_{\phantom{i}\,\,i}=\omega^i_{\phantom{0}0} 
    \thickspace(i=1,2,3)\thickspace \text{and}\thickspace
    \omega^i_{\phantom{j}j}=-\omega^j_{\phantom{i}\,\,i}
    \thickspace (i,j=1,2,3, i\ne j)
\]
and are related to the basis $\theta^i$ by
\[
    d\theta^i=-\omega^i_{\phantom{k}k}\wedge\theta^k.
\]
The solution of this system is found to be
\[
   \omega^0_{\phantom{1}1}=
        -\frac{\partial}{\partial r}\lambda(r,t)e^{-\alpha(r,t)}\,\theta^0
           +\frac{\partial}{\partial t}\alpha(r,t)e^{\lambda(r,t)}\,\theta^1,
\]
\[
   \omega^0_{\phantom{2}2}=\omega^0_{\phantom{3}3}=\omega^2_{\phantom{0}0}
    =\omega^3_{\phantom{0}0}=0,
\]
\[
    \omega^1_{\phantom{2}2}=-\frac{1}{r}e^{-\alpha(r,t)}\,\theta^2
    =-\omega^2_{\phantom{1}1},
\]
\[
    \omega^1_{\phantom{3}3}=-\frac{1}{r}e^{-\alpha(r,t)}
     \,\theta^3=-\omega^3_{\phantom{1}1},
\]
\[
     \omega^2_{\phantom{3}3}=-\frac{1}{r}\,(\text{cot}\,\theta)\,
     \theta^3=-\omega^3_{\phantom{2}2}.
\]

The Cartan structural equations then take on the form
\[
    \Omega^i_{\phantom{j}j}=d\omega^i_{\phantom{j}j}
    +\omega^i_{\phantom{k}k}\wedge\omega^k_{\phantom{k}j}.
\]

By direct calculation,
\begin{multline*}
   \Omega^0_{\phantom{0}1}=\\e^{2\lambda(r,t)}
    \left[\frac{\partial^2}{\partial t^2}\alpha(r,t)
    +\frac{\partial}{\partial t}\lambda(r,t)\frac{\partial}{\partial t}
     \alpha(r,t)
     +\left(\frac{\partial}{\partial t}\alpha(r,t)\right)^2\right]\,
     \theta^0\wedge\theta^1\\
  +e^{-2\alpha(r,t)}\left[\frac{\partial^2}{\partial r^2}\lambda(r,t)
     -\frac{\partial}{\partial r}\lambda(r,t)
      \frac{\partial}{\partial r}\alpha(r,t)
      -\left(\frac{\partial}{\partial r}\lambda(r,t)\right)^2
      \right]\,\theta^0\wedge\theta^1,
\end{multline*}
\begin{equation*}
    \Omega^0_{\phantom{2}2}=
    \frac{1}{r}\frac{\partial}{\partial r}\lambda(r,t)
       e^{-2\alpha(r,t)}\,\theta^0\wedge\theta^2
         -\frac{1}{r}\frac{\partial}{\partial t}\alpha(r,t)
          e^{\lambda(r,t)}e^{-\alpha(r,t)}\,\theta^1\wedge\theta^2,
\end{equation*}
\begin{equation*}
   \Omega^0_{\phantom{0}3}=
    \frac{1}{r}\frac{\partial}{\partial r}\lambda(r,t)
    e^{-2\alpha(r,t)}\,\theta^0\wedge\theta^3
      -\frac{1}{r}\frac{\partial}{\partial t}\alpha(r,t)
         e^{\lambda(r,t)}e^{-\alpha(r,t)}\,\theta^1\wedge\theta^3,
\end{equation*}
\begin{equation*}
   \Omega^1_{\phantom{2}2}=
   \frac{1}{r}e^{-\alpha(r,t)}e^{\lambda(r,t)}
      \frac{\partial}{\partial t}\alpha(r,t)\,\theta^0\wedge\theta^2
         +\frac{1}{r}e^{-2\alpha(r,t)}\frac{\partial}{\partial r}
             \alpha(r,t)\,\theta^1\wedge\theta^2,
\end{equation*}
\begin{equation*}
    \Omega^1_{\phantom{3}3}=
     \frac{1}{r}\frac{\partial}{\partial t}\alpha(r,t)
      e^{-\alpha(r,t)}e^{\lambda(r,t)}\,\theta^0\wedge\theta^3
         +\frac{1}{r}\frac{\partial}{\partial r}\alpha(r,t)
          e^{-2\alpha(r,t)}\,\theta^1\wedge\theta^3,
\end{equation*}
\begin{equation*}
   \Omega^2_{\phantom{3}3}=
      \frac{1}{r^2}\left(1-e^{-2\alpha(r,t)}\right)\,\theta^2\wedge\theta^3.
\end{equation*}

The final step is to read off the components of the Riemann curvature
tensor from the formula
\[
    \Omega^i_{\phantom{j}j}=-\frac{1}{2}R_{mnj}^{\phantom{mnj}i}
    \,\theta^m\wedge\theta^n.
\]
Following Hughston and Tod, we omit the hats, even though the
components are in the orthonormal frame.
(The negative sign is needed because Hughston and Tod use different sign
conventions in defining the curvature tensor.)

\textbf{Note:}\thickspace\emph{all other components are zero}.
\begin{multline*}
   R_{011}^{\phantom{000}0}=-e^{2\lambda(r,t)}
   \left[\frac{\partial^2}{\partial t^2}\alpha(r,t)
   +\frac{\partial}{\partial t}\lambda(r,t)
    \frac{\partial}{\partial t}\alpha(r,t)
      +\left(\frac{\partial}{\partial t}\alpha(r,t)\right)^2
         \right]\\
  -e^{-2\alpha(r,t)}\left[\frac{\partial^2}{\partial r^2}\lambda(r,t)
    -\frac{\partial}{\partial r}\lambda(r,t)
       \frac{\partial}{\partial r}\alpha(r,t)
       -\left(\frac{\partial}{\partial r}\lambda(r,t)\right)^2
         \right],
\end{multline*}
\begin{equation*}
   R_{022}^{\phantom{000}0}=
    -\frac{1}{r}e^{-2\alpha(r,t)}\frac{\partial}{\partial r}
     \lambda(r,t)=R_{033}^{\phantom{333}0},
\end{equation*}
\begin{equation}\label{E:Riem1}
    R_{122}^{\phantom{000}1}=
    -\frac{1}{r}e^{-2\alpha(r,t)}\frac{\partial}{\partial r}
     \alpha(r,t)=R_{133}^{\phantom{333}1},
\end{equation}
\begin{equation*}
    R_{233}^{\phantom{000}2}=
    -\frac{1}{r^2}\left(1-e^{-2\alpha(r,t)}\right),
\end{equation*}
\begin{equation*}
    R_{122}^{\phantom{000}0}=
     \frac{1}{r}e^{\lambda(r,t)}e^{-\alpha(r,t)}
        \frac{\partial}{\partial t}\alpha(r,t)=R_{133}^{\phantom{000}0}.
\end{equation*}

The components of the Einstein tensor are listed next, including the
new component, $G_{\hat{t}\hat{r}}$.
\begin{equation*}
   G_{\hat{t}\hat{t}}=
    \frac{2}{r}e^{-2\alpha(r,t)}\frac{\partial}{\partial r}
     \alpha(r,t)
           +\frac{1}{r^2}\left(1-e^{-2\alpha(r,t)}\right),
\end{equation*}
\begin{equation*}
   G_{\hat{r}\hat{r}}=
    -\frac{2}{r}e^{-2\alpha(r,t)}\frac{\partial}{\partial r}
     \lambda(r,t)-\frac{1}{r^2}
        \left(1-e^{-2\alpha(r,t)}\right),
\end{equation*}
\begin{equation}\label{E:Einstein2}
   G_{\hat{t}\hat{r}}=
     \frac{2}{r}e^{\lambda(r,t)}e^{-\alpha(r,t)}
         \frac{\partial}{\partial t}\alpha(r,t),
\end{equation}
\begin{multline*}
   G_{\hat{\theta}\hat{\theta}}=G_{\hat{\phi}\hat{\phi}}\\
   =-e^{2\lambda(r,t)}\left[\frac{\partial^2}{\partial t^2}
    \alpha(r,t)+\frac{\partial}{\partial t}\lambda(r,t)
    \frac{\partial}{\partial t}\alpha(r,t)
       +\left(\frac{\partial}{\partial t}\alpha(r,t)\right)^2
          \right]\\
    -e^{-2\alpha(r,t)}\left[\frac{\partial^2}{\partial r^2}
     \lambda(r,t)-\frac{\partial}{\partial r}\lambda(r,t)
      \frac{\partial}{\partial r}\alpha(r,t)
         -\left(\frac{\partial}{\partial r}\lambda(r,t)\right)^2
            \right]\\ 
        -\frac{1}{r}e^{-2\alpha(r,t)}\left(\frac{\partial}{\partial r}
            \lambda(r,t)+\frac{\partial}{\partial r}\alpha(r,t)\right).         
 \end{multline*}

\textbf{Remark:}  It is assumed that the results in this section 
are of interest in their own right.  To ensure accuracy, the calculations
were checked using the traditional method for determining the
components of the Riemann curvature tensor.

\subsection{WEC Violation}

According to Roman~\cite{tR93}, $T_{\hat{t}\hat{r}}$ is interpreted as energy 
flux, i.e., $T_{\hat{t}\hat{r}}=\pm f$, where $f$ is the energy flux in the 
outward radial direction.  The sign depends on whether the wormhole is
attractive or repulsive.  The expression for
$T_{\hat{\alpha}\hat{\beta}}\mu^{\hat{\alpha}}\mu^{\hat{\beta}}$ now becomes
\begin{multline}\label{E:exotic4}
   \rho c^2-\tau\pm 2f\\
     =\frac{1}{8\pi Gc^{-4}}\left[\frac{2}{r}e^{-2\alpha(r,t)}
      \left(\frac{\partial}{\partial r}\alpha(r,t)
      -\frac{\partial}{\partial r}
             \lambda(r,t)\right)\right.\\
    \left.\pm\frac{4}{r}e^{\lambda(r,t)}e^{-\alpha(r,t)}
        \frac{\partial}{\partial t}\alpha(r,t)\right].
\end{multline}
As expected, if the shape function is independent of time, then
Eq.~(\ref{E:exotic4}) reduces to Eq.~(\ref{E:exotic3}).  Otherwise,
if the last term is positive (resp. negative), the energy violation
is less severe (resp. more severe).  It is unlikely, however, that
the energy violation can be eliminated completely, as will be
seen below.

The expression for
\begin{equation}\label{E:exotic5}
   T_{\hat{t}\hat{t}}+T_{\hat{\theta}\hat{\theta}}
    =T_{\hat{t}\hat{t}}+T_{\hat{\phi}\hat{\phi}}
    =\rho c^2+p
\end{equation}
is also available from Eq.~(\ref{E:Einstein2}).  Although of
interest, the results are complicated and difficult to analyze
in the absence of specific functions.

\subsection{Traversability Conditions}
For the time-dependent case an analysis of the tidal gravitational
forces is somewhat more complicated.
From Eq.~(\ref{E:Riem1}) 
   \begin{multline}\label{E:lateraltidal2}
   \left|R_{\hat{2}'\hat{0}'\hat{2}'\hat{0}'}\right|
   =\left|R_{\hat{3}'\hat{0}'\hat{3}'\hat{0}'}\right|\\
   =\gamma^2\left|R_{\hat{\theta}\hat{t}\hat{\theta}\hat{t}}\right|
   +\gamma^2\left(\frac{v}{c}\right)^2
      \left|R_{\hat{\theta}\hat{r}\hat{\theta}\hat{r}}\right|
      +2\gamma^2\left(\frac{v}{c}\right)
      \left|R_{\hat{\theta}\hat{t}\hat{\theta}\hat{r}}\right|\\
   =\gamma^2\left|\frac{1}{r}\frac{\partial}{\partial r}
      \lambda(r,t)e^{-2\alpha(r,t)}\right|
        +\gamma^2\left(\frac{v}{c}\right)^2
         \left|\frac{1}{r}\frac{\partial}{\partial r}\alpha(r,t)
               e^{-2\alpha(r,t)}\right|\\
   +2\gamma^2\left(\frac{v}{c}\right)\left|\frac{1}{r}
       \frac{\partial}{\partial t}\alpha(r,t)
           e^{\lambda(r,t)}e^{-\alpha(r,t)}\right|.
\end{multline} 
The resulting constraint complements Eq.~(\ref{E:exotic4}): whenever
$\alpha(r,t)$ increases or decreases fast enough, the lateral tidal force increases
in time.  So while Eq.~(\ref{E:exotic4}) suggests
that the energy violation can be much reduced or even eliminated,
the increasing tidal force is likely to make the wormhole nontraversable.

Similar comments apply to the radial tidal constraint involving 
\[
  \left|R_{\hat{1}'\hat{0}'\hat{1}'\hat{0}'}\right|=
  \left|R_{\hat{r}\hat{t}\hat{r}\hat{t}}\right|.
\]
One can see from Eq.~(\ref{E:Riem1}) that if $\alpha(r,t)$
changes fast enough, then the tidal force may very well increase
in time.  For example, in the special case where $\lambda$ is constant,
even a small increase in the time rate of change of $\alpha$
(or a steady \emph{decrease})
will produce an ever-increasing tidal force.

As a final remark, to see how an observer's four-acceleration varies with
time, let $V^\mu=dx^\mu/d\tau=(e^{\lambda(r,t)},0,0,0)$ be the four-velocity
of an observer who is at rest with respect to the $r, \theta, \phi$
coordinate system.  Then his or her four-acceleration is
\[
   a^\mu=V^\mu_{\phantom{\mu}\,;\,\nu}V^\nu=
   \left(V^\mu_{\phantom{\mu}\,,\,\nu}+\Gamma_{\alpha\nu}^\mu V^\alpha\right)V^\nu.
\]
For the line element in Eq.~(\ref{E:line3}), $a^t=0$, as one would expect,
while the radial component is
\begin{equation*}
a^r=\Gamma_{tt}^r\left(\frac{dt}{d\tau}\right)^2=-c^2e^{-2\alpha(r,t)}\frac{\partial}{\partial r}
\lambda(r,t).
\end{equation*}


\section{Conclusion}
We have shown that for wormholes to meet the quantum inequality constraints due to Ford
and Roman~ \cite{lF96}, the shape function can have the generic form
$b=r(1-e^{-2\alpha(r)})$, but the slope of $\alpha$ is subject to some restrictions.
The redshift function may also be quite general.  Violations of the weak
energy condition were discussed in terms of these functions.  Analogous solutions were 
obtained for the corresponding time-dependent functions.  It turned
out that the energy violations may vary but cannot be completely 
eliminated, even temporarily.
  
  Specific examples of the shape and
redshift functions have shown that the resulting wormhole is traversable for humanoid
travelers.  The conclusions are valid for the MTY wormhole.  While initially promising,
the MTY wormhole, as described in Ref. \cite{MTY88}, requires a huge apparatus that does
not appear to be technically feasible in the forseeable future.

On the other hand, if a way can be found to spread the exotic material over a larger
region, or if, under certain conditions, the quantum inequality can be relaxed
somewhat, then the solution will be closer to being realizable.  Until (if ever)
this happens, the solution presented here may be the best that can be done.

\vspace{16pt}

\begin{center}
    APPENDIX
\end{center}
  
The calculations in the Appendix refer to the line element ~(\ref{E:line1}) with
$\gamma(r)=-\lambda(r,t)$.

A partial list of nonzero Christoffel symbols and Riemann curvature components are
 listed next.  (\emph{Reminder:} to obtain Eq. ~(\ref{E:Riemann}), replace $\lambda(r,t)$
by $\alpha(r)$ and $\alpha$ by $\alpha(r-r_0)$.)
\begin{align*}
  \Gamma_{rt}^{t}&=-\frac{\partial}{\partial r}
     \lambda(r,t)  &
               \Gamma_{\theta\theta}^{r}&=-re^{-2\alpha} 
                      &\Gamma_{\phi\phi}^{r}&=-re^{-2\alpha}
               \text{sin}^2\theta\\
               \Gamma_{rr}^{r}&=\alpha' &  
               \Gamma_{r\theta}^{\theta}
                 &=\frac{1}{r}  &
                   \Gamma_{\phi\phi}^{\theta}&=-\text{sin}\,\theta\,
                   \text{cos}\,\theta\\
                   \Gamma_{\phi\theta}^{\phi}&=\text{cot}\,\theta
                       &\Gamma^t_{tt}&=-\frac{\partial}{\partial t}\lambda(r,t)
\end{align*}
\begin{equation*}
   R_{rtr}^{t}=-e^{2\alpha}e^{2\lambda(r,t)}R_{trt}^{r}
     =\frac{\partial^2}{\partial r^2}
      \lambda(r,t)-\alpha'
     \frac{\partial}{\partial r}\lambda(r,t)
          -\left(\frac{\partial}{\partial r}
           \lambda(r,t)\right)^2
\end{equation*}
\begin{equation*}
  R_{\theta t\theta}^{t}=-r^2e^{2\lambda(r,t)}
   R_{t\theta t}^{\theta}
  =re^{-2\alpha}\frac
   {\partial}{\partial r}\lambda(r,t)
\end{equation*}
\begin{equation*}
   R_{\phi t\phi}^{t}=-r^2e^{2\lambda(r,t)}
    \,\text{sin}^2\theta 
   \,R_{t\phi t}^{\phi}
    =re^{-2\alpha}\,\text{sin}^2\theta\,
     \frac{\partial}{\partial r}
    \lambda(r,t)
\end{equation*}
\begin{equation*}
  R_{\theta r\theta}^{r}=r^2e^{-2\alpha}
   R_{r\theta r}^{\theta}
   =re^{-2\alpha}\alpha'
\end{equation*}
\begin{equation*}
   R_{\phi r\phi}^{r}=r^2e^{-2\alpha}\,\text{sin}^2\theta\,
    R_{r\phi r}^{\phi}
   =\text{sin}^2\theta\,\left(
    re^{-2\alpha}\alpha'\right)
\end{equation*}
\begin{equation*}
  R_{\phi\theta\phi}^{\theta}=\text{sin}^2\theta\,
  R_{\theta\phi\theta}^{\phi}=
    \text{sin}^2\theta\,(1-e^{-2\alpha})
\end{equation*}

The components of the Ricci tensor in the orthonormal frame are given next.
(\emph{Reminder:} to obtain Eq.~(\ref{E:Einstein}), replace $\lambda(r,t)$ 
by $\alpha(r)$ and $\alpha$ by $\alpha(r-r_0)$.)
\begin{multline*}
   R_{\hat{t}\hat{t}}=-e^{-2\alpha}\left[
   \frac{\partial^2}{\partial r^2}
   \lambda(r,t)-\alpha'
   \frac{\partial}{\partial r}\lambda(r,t)
   -\left(\frac{\partial}{\partial r}
   \lambda(r,t)\right)^2\right]\\
        -\frac{2}{r}e^{-2\alpha}\frac{\partial}{\partial r}\lambda(r,t)
\end{multline*}
\begin{multline*}
   R_{\hat{r}\hat{r}}=e^{-2\alpha}\left[
   \frac{\partial^2}{\partial r^2}\lambda(r,t)
     -\alpha'\frac{\partial}{\partial r}\lambda(r,t)
          -\left(\frac{\partial}{\partial r}
    \lambda(r,t)\right)^2\right]
    +\frac{2}{r}e^{-2\alpha}\alpha'
\end{multline*}
\begin{equation*}
   R_{\hat{\theta}\hat{\theta}}=R_{\hat{\phi}\hat{\phi}}
    =\frac{1}{r}e^{-2\alpha}\frac{\partial}{\partial r}\lambda(r,t)
   +\frac{1}{r}e^{-2\alpha}\alpha'
       +\frac{1}{r^2}(1-e^{-2\alpha})
\end{equation*}

E-mail address: kuhfitti@msoe.edu

\end{document}